# On the Mutual Coefficient of Restitution in Two Car Collinear Collisions

Milan Batista

University of Ljubljana, Faculty of Maritime Studies and Transportation

Pot pomorscakov 4, Slovenia, EU

milan.batista@fpp.edu

(Updated Feb. 2006)

**Abstract**

In the paper two car collinear collisions are discussed using Newton's law of mechanics, conservation of energy and linear constitutive law connecting impact force and crush. Two ways of calculating the mutual restitution coefficient are given: one based on car masses and one based on car stiffness. A numerical example of an actual test is provided.

## 1. Introduction

For the modeling of the collinear car collision two methods are usually used. The first is the so-called impulse-momentum method based on classical Poisson impact theory, which replaces the forces with the impulses ([3], [11]). The second method treats a car as a deformable body; so the constitutive law connecting contact force with crush is necessary. For the compression phase of impact the linear model of force is usually adopted and the models differ in the way the restitution phase of collision is treated ([7], [13], [14], [17]).

The purpose of this paper is to extend the linear force model discussed in [1] to the collinear impact of two cars. In the quoted article it is proposed that a car is characterized by its mass, stiffness and limit velocity for permanent crush. The latter properties can be established by a fixed barrier crush test. Also, the proposed restitution model is simple: rebound velocity is constant. The question arises as to how these characteristics can be incorporated into the two car collision model since it is well known that the mutual coefficient of restitution is the characteristic of impact; i.e., it is a two car system and not the property of an individual car ([2], [17]).

To answer the above question, first the well-known theory of central impact is specialized for collinear car collisions. The kinetic energy losses are then discussed and the restitution coefficient is related to them. The third section of the paper discusses two models for calculating the mutual restitution coefficient based on individual car characteristics. The last section is devoted to a description of the use of the present theory in accident reconstruction practice. The section ends with a numerical example.

**2. Two car collinear collision**

Consider a collinear impact between two cars where collinear impact refers to rear-end and head-on collisions. Before impact the cars have velocities $v_1$ and $v_2$ respectively and after impact they have velocities $u_1$ and $u_2$ (Figure 1).

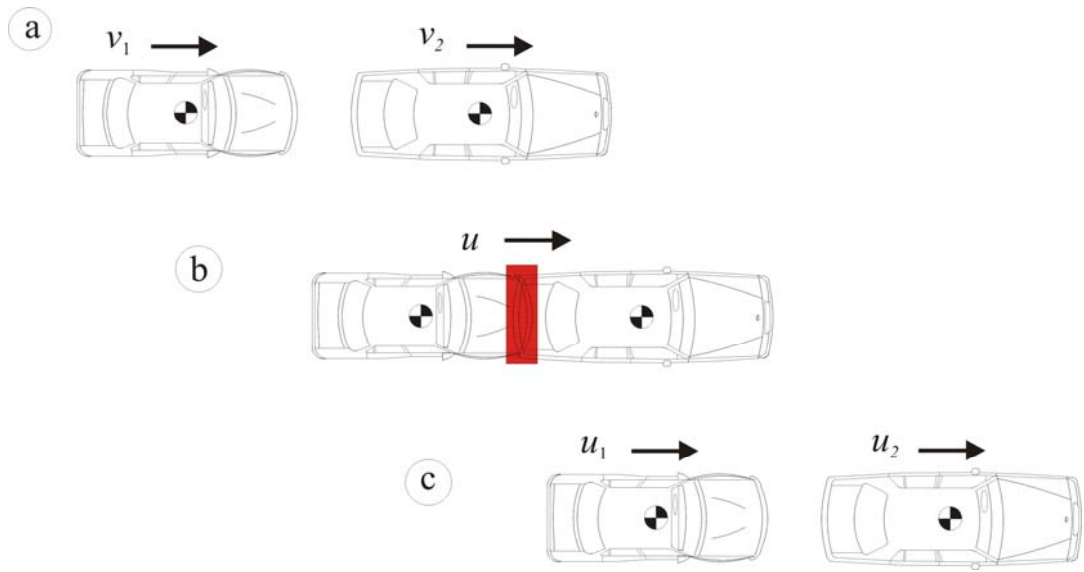

**Figure 1.** The two car impact: (a) pre-impact velocities, (b) end of compression velocity, (c) post-impact velocities

In the collision phase the movement of cars is governed by Newton's 2nd and 3rd laws (Figure 2). On the basis of these laws equations of motion of the cars can be written as follows

$$m_1 \frac{dv_1}{dt} = -F \quad \text{and} \quad m_2 \frac{dv_2}{dt} = F \tag{1}$$

where $m_1$ and $m_2$ are the masses of the cars and $F$ is contact force.

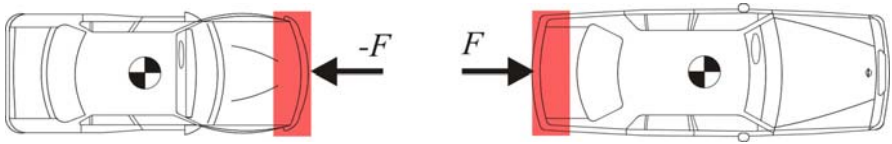

**Figure 2.** Newton's 3rd law applied to collinear impact of two cars

Following Poisson's hypothesis ([16]), the impact is divided into two phases: compression and restitution. In the compression phase the contact force $F$ raises and the cars are deformed. The compression phase terminates when the relative velocity of cars vanishes; i.e., when cars have equal velocity (Figure 1). The compression phase (1) thus integrates the changes from initial velocities to common velocity $u$. This leads to the following system of equations

$$m_1(u - v_1) = -P_c \quad m_2(u - v_2) = P_c \tag{2}$$

where $P_c \equiv \int_0^{\tau_c} F \, dt$ is compression impulse and $\tau_c$ compression time. From (2) one obtains the velocity after compression

$$\boxed{u = \frac{m_1 v_1 + m_2 v_2}{m_1 + m_2}} \tag{3}$$

and the compression impulse

$$P_c = \frac{m_1 m_2}{m_1 + m_2}(v_1 - v_2) \tag{4}$$

In the restitution phase the elastic part of internal energy is released. Equations (1) are integrated from $u$ to the end velocities, which gives two equations for three unknowns

$$m_1(u_1 - u) = -P_r \qquad m_2(u_2 - u) = P_r \tag{5}$$

where $P_r \equiv \int_0^{\tau_c} F\,dt$ is restitution impulse and $\tau_r$ is restitution time. In order to solve system (5) for an unknown's post-impact velocity and restitution impulse the constitutive equation is needed. According to the Poisson hypothesis the restitution impulse is proportional to compression impulse

$$P_r = eP_c \tag{6}$$

where $e$ is the restitution coefficient. Because contact force is non-negative, so are compression and restitution impulse. From (6) this implies that $e \geq 0$.

**Note.** Instead of (6), one can use Newton's kinematical definition of restitution coefficient

$$e = \frac{u_2 - u_1}{v_1 - v_2}$$

which is in the case of centric impact without friction equivalent to Poisson's definition. However in the case of non-centric impact with friction Newton's model could lead to overall energy increase ([12]).

The total impulse is $P = P_c + P_r$ so by using (4) and (6)

$$P = (1+e)\frac{m_1 m_2}{m_1 + m_2}\Delta v \qquad (7)$$

Solving (5) and (6) and taking into account (4) gives the well known formulas (see for example [3], [11]) for the cars post-impact velocities

$$\boxed{\begin{aligned} u_1 &= u - e\frac{m_2}{m_1 + m_2}\Delta v = v_1 - \frac{(1+e)m_2}{m_1 + m_2}\Delta v \\ u_2 &= u + e\frac{m_1}{m_1 + m_2}\Delta v = v_2 + \frac{(1+e)m_1}{m_1 + m_2}\Delta v \end{aligned}} \qquad (8)$$

where $\Delta v = v_1 - v_2$. The above equations can be used for calculation of post-impact velocities if pre-impact velocities are known, masses of cars are known and, in addition, the restitution coefficient is known.

**3. Energy consideration**

At car impact the kinetic energy is dissipated. Applying the principle of conservation of energy one obtains, after compression,

$$\frac{m_1 v_1^2}{2} + \frac{m_2 v_2^2}{2} = \frac{(m_1 + m_2)u^2}{2} + \Delta E_m \qquad (9)$$

where $\Delta E_m$ is maximal kinetic energy lost (or maximal energy absorbed by crush). By using (3) one has

$$\Delta E_m = \frac{1}{2}\frac{m_1 m_2}{m_1 + m_2}\Delta v^2 \qquad (10)$$

Similarly, by applying the principle of conservation of energy to the overall impact process

$$\frac{m_1 v_1^2}{2} + \frac{m_1 v_1^2}{2} = \frac{m_1 u_1^2}{2} + \frac{m_1 u_1^2}{2} + \Delta E \qquad (11)$$

one finds the well known formula for total kinetic energy lost (see for example [11])

$$\Delta E = \frac{1}{2}\left(1 - e^2\right) \frac{m_1 m_2}{m_1 + m_2} \Delta v^2 \qquad (12)$$

Since, by the law of thermodynamics, $\Delta E \geq 0$, it follows from (12) that $e \leq 1$. Now, from (10) and (12) one has $\Delta E = \left(1 - e^2\right) \Delta E_m$, so the mutual restitution coefficient is given by ([11])

$$\boxed{e = \sqrt{1 - \frac{\Delta E}{\Delta E_a}} = \sqrt{\frac{\Delta E_0}{\Delta E_m}}} \qquad (13)$$

where $\Delta E_0 \equiv \Delta E_m - \Delta E$ is the rebound energy. The formula obtained is the basis for relating the mutual coefficient of restitution $e$ with the restitution coefficients obtained for individual cars in the fixed barrier test.

**4. The mutual coefficient of restitution**

Let $v_{T1}$ be a barrier test velocity of a first car and $v_{T2}$ a barrier test velocity of a second car. Let these velocities be such that the maximal kinetic energy lost can be written as

$$\Delta E_m = \frac{m_1 v_{T1}^2}{2} + \frac{m_2 v_{T2}^2}{2} \qquad (14)$$

and in addition the rebound energy can be written as (see [9])

$$\Delta E_0 = \frac{m_1 e_1^2 v_{T1}^2}{2} + \frac{m_2 e_2^2 v_{T2}^2}{2} \qquad (15)$$

The mutual restitution coefficient is therefore from (13), (14) and (15), by using (10),

$$e = \sqrt{\frac{m_1 e_1^2 v_{T1}^2 + m_2 e_2^2 v_{T2}^2}{m_1 v_{T1}^2 + m_2 v_{T2}^2}} \qquad (16)$$

For the model of the barrier test proposed in [1] the restitution coefficients of cars are

$$e_1 = \min\left(1, \frac{v_{01}}{v_{T1}}\right) \quad \text{and} \quad e_2 = \min\left(1, \frac{v_{02}}{v_{T2}}\right) \qquad (17)$$

where $v_{01}$ and $v_{02}$ are limited impact velocities where all the crush is recoverable ([1]). The task is now to determine appropriate test velocities of cars which satisfy (14).

4. 1 *Model A - stiffness based mutual restitution coefficient.*

Let $v_{T1}$ be the barrier test velocity (or barrier equivalent velocity [8]) of the first car for the same crush as in a two car impact and $v_{T2}$ the barrier test velocity for the same crush for the second car. Then the test velocities for the same crush must satisfy relations ([1], [8])

$$\frac{m_1 v_{T1}^2}{2} = \frac{k_1 \delta_{m1}^2}{2} \quad \text{and} \quad \frac{m_2 v_{T2}^2}{2} = \frac{k_2 \delta_{m2}^2}{2} \qquad (18)$$

where $k_1$ and $k_2$ are stiffness of the cars and $\delta_{m1}$ and $\delta_{m2}$ are actual maximal dynamics crush of the cars. From (18) one has

$$v_{T1} = \sqrt{\frac{k_1}{m_1}}\delta_{m1} \quad \text{and} \quad v_{T2} = \sqrt{\frac{k_2}{m_2}}\delta_{m2} \tag{19}$$

On the other hand, from (10), (14) and (18) it follows that

$$\Delta E_m = \frac{1}{2}\frac{m_1 m_2}{m_1 + m_2}\Delta v^2 = \frac{k_1 \delta_{m1}^2}{2} + \frac{k_2 \delta_{m2}^2}{2} \tag{20}$$

Defining overall maximal crush $\delta_m \equiv \delta_{m1} + \delta_{m2}$ and taking into account the law of action and reaction $k_1 \delta_{m1} = k_2 \delta_{m2}$ one obtains

$$\delta_{m1} = \frac{k_2}{k_1 + k_2}\delta_m \qquad \delta_{m2} = \frac{k_1}{k_1 + k_2}\delta_m \tag{21}$$

Substituting (21) into (20) yields

$$\Delta E_m = \frac{m\Delta v^2}{2} = \frac{k\delta_m^2}{2} \tag{22}$$

where $m$ is system mass and $k$ is system stiffness, given by

$$m \equiv \frac{m_1 m_2}{m_1 + m_2} \qquad k \equiv \frac{k_1 k_2}{k_1 + k_2} \tag{23}$$

From (22) one has $\delta_m = \sqrt{\frac{m}{k}}|\Delta v|$ and therefore from (19) the required test velocities are (see also [8])

$$\boxed{v_{T1} = \sqrt{\frac{k}{k_1}\frac{m}{m_1}}|\Delta v| \quad \text{and} \quad v_{T2} = \sqrt{\frac{k}{k_2}\frac{m}{m_2}}|\Delta v|} \tag{24}$$

Substituting (24) into (14) leads to identity $\frac{1}{k} = \frac{1}{k_1} + \frac{1}{k_2}$ and substituting it into (16) provides the required mutual restitution coefficient

$$e = \sqrt{\frac{k_2 e_1^2 + k_1 e_2^2}{k_1 + k_2}} \qquad (25)$$

This equation for the calculation of *e* were published by various authors ([4],[5],[15]). Knowing the mass and stiffness of the cars and $\Delta v$ one can calculate test velocities from (24), restitution of individual cars from (17), the mutual restitution coefficient from (25) and post-impact velocities from (8).

*4. 2 Model B - mass based mutual restitution coefficient.*

This model does not include cars' stiffness and it's based on (10) and (14) only. Equating (10) and (14) results in the equation

$$m \Delta v^2 = m_1 v_{T1}^2 + m_2 v_{T2}^2 \qquad (26)$$

for two unknowns. To solve it one could set

$$v_{T1} = v_1 - v_0 \qquad v_{T2} = v_2 - v_0 \qquad (27)$$

where $v_0$ is a new unknown velocity. Substituting (27) into (14) one obtains after simplification $\left[ m_1 (v_1 - v_0) + m_2 (v_2 - v_0) \right]^2 = 0$, so

$$v_0 = \frac{m_1 v_1 + m_2 v_2}{m_1 + m_2} \qquad (28)$$

This is in fact the velocity of the centre of the mass of colliding cars. Substituting (28) into (27) yields unknown test velocities

$$\boxed{v_{T1} = \frac{m_2(v_1 - v_2)}{m_1 + m_2} \qquad v_{T2} = -\frac{m_1(v_1 - v_2)}{m_1 + m_2}} \tag{29}$$

Note that in calculation of restitution coefficients (17) the absolute values of test velocities should be used. Substituting (29) into (16) gives the mutual restitution coefficient

$$\boxed{e = \sqrt{\frac{m_2 e_1^2 + m_1 e_2^2}{m_1 + m_2}}} \tag{30}$$

This formula was derived by different arguments of Howard et al ([9]) and is also quoted by Watts et al ([18]).

4.3 *Compartment of the models*

Comparing (24) and (25) one finds that test velocities of both models are the same if stiffness is proportional to the mass; i.e., $k_1 = k_0 m_1$ and $k_2 = k_0 m_2$ where $k_0$ is a constant.

While the test velocities of the models differ, the mutual restitution coefficient differs only in the case when just one car is crushed permanently, since

- when $v_{T1} \leq v_{01}$ and $v_{T2} \leq v_{02}$ then both $e_1 = e_2 = 1$ so by (25) or (30) it follows $e = 1$ and
- when $v_{T1} > v_{01}$ and $v_{T2} > v_{02}$ then substituting (17) and appropriate test velocities into (25) or (30), and taking (10) into account, yields

$$e = \sqrt{\frac{m_1 v_{01}^2 + m_2 v_{02}^2}{m \Delta v^2}} \tag{31}$$

Note that (31) can not be used directly for calculating the mutual restitution coefficient in advance since the classification of impact--fully elastic, fully plastic or mixed--depends on test velocities.

At last the question arises as to which model is more physically justified. While Model A has a sound physical base connecting test velocities with crushes, Model B requires some additional analysis. It turns out that it can be interpreted as follows. The compression impulse (4), can be written by using $(23)_1$ as $P_c = m\Delta v$. Using (2) one could define test velocities of individual cars as velocities resulting at the end of the compression phase in a fixed barrier test as the same impulse as in an actual two car collision; i.e.,

$$P_c = m|\Delta v| = m_1 v_{T1} = m_2 v_{T2} \tag{32}$$

From this equation, test velocities given already by (29) result. Now by (6) restitution impulse is $P_r = eP_c = em|\Delta v|$, so by (5) and (32) one must have $em|\Delta v| = e_1 m_1 v_{T1} = e_2 m_2 v_{T2}$. But this can be fulfilled only in the special case when $e_1 = e_2$, and consequently, by (30), when $e = e_1$. This consequence raises a doubt about Model B's adequacy for general use.

4.4 *Examples*

The above formulas were implemented into the spreadsheet program (Table 1).As the example, a full scale test (test no. 7) reported by Cipriani et al ([6]) was executed. In this test the bullet car made impact with the rear of the target car at a velocity of 5 m/s or 18 km/h. The mass of the cars and their stiffness was taken from the report; however, the limit speed was taken to be 4 km/h for both cars ([1]). The result of the calculation is shown in Table 2. The calculated velocity difference for the target car is 14.8 km/h, which differs from that measured (3.9 m/s or 14.0 km/h) by about 5%. The calculated velocity change for the bullet car is 11.3 km/h and the measured one was 2.9 m/s or 10.4 km/h. The discrepancy is thus about 7%. If one takes the limit speed to be 3 km/h, then the calculated value of velocity change for the bullet car is 13.6 km/h, differing from that measured by about 2%, and the calculated value of velocity change for the target car is 10.4, which actually matches the measured value.
.

**Table 1.** Spreadsheet program for calculation of post-impact velocities
Full scale test 7 of Cipriani et al ([6])

|  |  | Vehicle 1 |  | Vehicle 2 |
|---|---|---|---|---|
| mass | kg | 1146 |  | 1495 |
| stiffness | kN/m | 886.07 |  | 1564.687 |
| limit velocity | km/h | 4 |  | 4 |
| impact velocity | km/h | 18 |  | 0 |
| Delta V | km/h |  | 18.00 |  |
| velocity after compression | km/h |  | 7.81 |  |
| system mass | kg |  | 648.72 |  |
| system stiffness | kN/m |  | 565.71 |  |
| test velocity | km/h | 10.82 |  | 7.13 |
| test restitution |  | 0.37 |  | 0.56 |
| restitution |  |  | 0.45 |  |
| post impact velocity | km/h | 3.24 |  | 11.31 |
| Delta V | km/h | 14.76 |  | -11.31 |
| Maximal crush | m | 0.11 |  | 0.06 |
| Residual crush | m | 0.07 |  | 0.03 |

**5. Accident Reconstruction**

In a real car accident the problem is not to determine post-impact velocities but usually the opposite; i.e., to calculate the pre-impact velocities. For determining pre-impact velocities, however, the post-impact velocities determined from skid-marks should be known. If only the permanent crushes of cars are known then only the velocity changes for individual cars in an accident can be calculated. If the characteristics of cars are known--i.e., mass, stiffness and limit velocity--then the problem is solved as follows. Let $\delta_{r1}$ be residual crush of the first vehicle. The maximal crush, then, is ([1])

$$\delta_{m1} = \delta_{r1} + \delta_{01} \tag{33}$$

where the recoverable part of crush is calculated as $\delta_{01} = v_{01}\sqrt{\dfrac{m_1}{k_1}}$. The maximal crush of the second car can be calculated in the same way or from Newton's 3rd law as

$$\delta_{m2} = \dfrac{k_1}{k_2}\delta_{m1} \tag{34}$$

The maximal energy lost at impact is then calculated from

$$\Delta E_m = \Delta E_{m1} + \Delta E_{m2} \tag{35}$$

where $\Delta E_{m1} = \dfrac{k_1 \delta_{m1}^2}{2}$ and $\Delta E_{m2} = \dfrac{k_2 \delta_{m2}^2}{2}$. The pre-impact velocity difference is thus, from (22),

$$\Delta v = \sqrt{\dfrac{2\Delta E_m}{m}} \tag{36}$$

To calculate velocity changes of individual vehicles the first test velocities are calculated by (18)

$$v_{T1} = \sqrt{\dfrac{2\Delta E_{m1}}{m_1}} \qquad v_{T2} = \sqrt{\dfrac{2\Delta E_{m2}}{m_2}} \tag{37}$$

From (17) the restitution coefficient for individual cars are calculated and from (25) the mutual coefficient of restitution. From (8) the velocity differences of individual cars at impact are

$$\Delta v_1 = v_1 - u_1 = \dfrac{(1+e)m_2}{m_1 + m_2}\Delta v \qquad \Delta v_2 = v_2 - u_2 = -\dfrac{(1+e)m_1}{m_1 + m_2}\Delta v \tag{38}$$

The above formulas were programmed into a spreadsheet program (Table 2). As the example, the car to car test described by Kerkhoff et al ([10]) is considered. In this test the test car (bullet) struck the rear of the stationary car (target) at a speed of 40.6 mph or 65 km/h. The actual measured $\Delta v$ was 22.6 mph or 36.2 km/h. As can be seen from Table 2, the calculated value $\Delta v_1$ for the bullet car is 36.1 km/h; i.e., the discrepancy between actual and calculated value is 0.2% and the calculated impact velocity 64.14 km/h differs from the actual by 1.3 %. Note that the deformation of the stationary car was not reported, so (34) is used for calculation of its maximal dynamic crush. The limit speed for both cars was taken to be 4 km/h ([1]). The discrepancy of calculated values in the previous case is so minimal because the actual low impact velocity tests were used for determination of stiffness. If one used for the calculation the default values of CRASH stiffness and appropriate calculated limit velocity for class 1 cars the discrepancy would increase. Thus, in this case the calculated velocity change of the bullet car is 38.5 km/h, which differs from the actual change by about 6% and the calculated $\Delta v$ is 52.2 km/h, differing by about 20%.

**Table 2.** Spreadsheet program for calculation of velocity differences at impact. Car to car test no 1 by Kerkhoff et al ([10])

|  |  |  | Vehicle 1 |  | Vehicle 2 |
|---|---|---|---|---|---|
| Data | mass | kg | 1100.44 |  | 1101.11 |
|  | stiffness | kN/m | 1681.91 |  | 872.89 |
|  | limit speed | km/h | 4.00 |  | 4.00 |
|  | crush | m | 0.16 |  | ? |
| recoverable crush |  | m | 0.03 |  | 0.04 |
| maximal crush |  | m | 0.19 |  | 0.36 |
| system mass |  | kg |  | 550.39 |  |
| system stiffness |  | kN/m |  | 574.65 |  |
| max energy lost |  | kJ | 29.86 |  | 57.53 |
| test velocity |  | km/h | 26.52 |  | 36.80 |
| test restitution |  |  | 0.15 |  | 0.11 |
| restitution |  |  |  | 0.12 |  |
| Delta V |  | km/h | 36.09 | 64.15 | -36.06 |